# Radio Frequency Energy Harvesting and Management for Wireless Sensor Networks

Adamu Murtala Zungeru, Li-Minn Ang, SRS. Prabaharan, Kah Phooi Seng

Department of Electrical and Electronics Engineering

The University of Nottingham

adamuzungeru@ieee.org

**ABSTRACT:** Radio Frequency (RF) Energy Harvesting holds a promising future for generating a small amount of electrical power to drive partial circuits in wirelessly communicating electronics devices. Reducing power consumption has become a major challenge in wireless sensor networks. As a vital factor affecting system cost and lifetime, energy consumption in wireless sensor networks is an emerging and active research area. This chapter presents a practical approach for RF Energy harvesting and management of the harvested and available energy for wireless sensor networks using the Improved Energy Efficient Ant Based Routing Algorithm (IEEABR) as our proposed algorithm. The chapter looks at measurement of the RF power density, calculation of the received power, storage of the harvested power, and management of the power in wireless sensor networks. The routing uses IEEABR technique for energy management. Practical and real-time implementations of the RF Energy using Powercast™ harvesters and simulations using the energy model of our Libelium Waspmote to verify the approach were performed. The chapter concludes with performance analysis of the harvested energy, comparison of IEEABR and other traditional energy management techniques, while also looking at open research areas of energy harvesting and management for wireless sensor networks.

## 1.0 INTRODUCTION

Finite electrical battery life is encouraging companies and researchers to come up with new ideas and technologies to drive wireless mobile devices for an enhanced period of time. Batteries add to size and their disposal adds to environmental pollution. For mobile and miniature electronics devices, a promising solution is available in capturing and storing the energy from external ambient sources, a technology known as energy harvesting. Other names for this type of technology are power harvesting, energy scavenging and free energy, which are derived from renewable energy [1]. In recent years the use of wireless devices is growing in many applications like mobile phones and sensor networks [2]. This increase in wireless applications has generated an increasing use of batteries. Many research teams are working on extending the battery life by reducing the consumption of the devices. Others teams have chosen to recycle ambient energy like in Micro-electromechanical Systems (MEMS) [3]. The charging of mobile devices is convenient because the user can do it easily, like for mobile phones. But for other applications, like wireless sensor nodes which are located in difficult to access environments, the charging of the batteries remain a major problem. This problem increases when the number of devices is large and are distributed in a wide area or located in inaccessible places. The research on RF energy harvesting provides reasonable techniques of overcoming these problems.

The rectification of microwave signals to DC power has been proposed and researched in the context of high-power beaming. It has been proposed for helicopter powering [4], solar power satellite [5], the SHARP System [6]. The DC power depends on the available RF power, the choice of antenna and frequency band. An Energy harvesting technique using electromagnetic energy specifically radio frequency is the focus of this chapter. Communication devices generally have Omni-directional antennas that propagate RF energy in most directions, which maximizes connectivity for mobile applications. The energy transmitted from the wireless sources is much higher up to 30W for 10GHz frequency [7], but only a small amount can be scavenged in the real environment. The rest is dissipated as heat or absorbed by other materials. RF power harvesting technique is also used in Radio Frequency Identification (RFID) tags and implantable electronics devices. Most commonly used wireless sensor nodes consume few µW in sleep mode and hundreds µW in active mode. A great factor contributing for energy harvesting research and development is ultra-low-power components.

The management of power available for sensor nodes as been dealt with to an extent using Ant Based Routing [8-16], which utilizes the behavior of real ants searching for food through pheromone deposition while dealing with problems that need to find paths to goals. The simulating behavior of ant colony leads to optimization of network parameters for the WSN routing process to provide maximum network lifetime.

The main goal of this chapter is to propose practical harvesting Radio Frequency Energy using Powercast Harvesters while managing the harvested and available energy of the sensor networks using our proposed Algorithm; Improved Energy Efficient Ant Based Routing, which help in the optimization of the available power. The objective is to power efficiently sensor networks with or without batteries to maintain network lifetime at a maximum without performance degradation.

The chapter is organized in the following format: The introductory part of the chapter provided in section 1, covers a general perspective and the objective of the chapter. Section 2 reviews energy harvesting systems and power consumption in WSNs. Section 3 gives detailed explanation of our radio frequency energy harvesting method using the Powercast harvesters. Section 4 looks in to the management of the harvested energy in our WSNs. Section 5 presents

Experimental set-up and results, while also looking at the simulation results and its environment. Finally, section 6 concludes the chapter with an open research problems and future work to be done, and a comparative summary of our results with the Energy-Efficient Ant-Based Routing (EEABR) algorithm, and Ad-hoc On-demand Distance Vector (AODV) which form strong energy management protocols.

## 2.0 REVIEW OF ENERGY HARVESTING SYSTEM AND POWER CONSUMPTION IN WSNs

For proper operation of sensor networks, a reliable energy harvesting techniques is needed. Over the years, so much work has been done on the research from both academic and industrial researchers on large scale energy from various renewable energy sources. Less attention has been paid to small scale energy harvesting techniques. Though, quite a number of works has been carried out on energy scavenging for WSNs. The Efficient far-field energy harvesting [17] uses a passively powered RF-DC conversion circuit operating at 906MHz to achieve power of up to 5.5μW. In a related work [18-21], all consider the little available RF energy while utilizing it to power the sensor networks. Bouchouicha et al [2] studied ambient RF energy harvesting in which two systems, the broad band without matching and narrow band were used to recover the RF energy. Among but not all of the available Energy harvesting system for Wireless sensors are; solar power, electromagnetic energy, thermal energy, wind energy, salinity gradients, kinetic energy, biomedical, piezoelectric, pyroelectric, thermoelectric, electrostatic, blood sugar, tree metabolic energy. These could be further classified in to three [22]; Thermal energy, Radiant energy, and Mechanical energy. Based on these, a table showing the comparison of the different and common energy scavenging techniques is as below in table 1.

**Table 1.0 Comparison of Energy Harvesting Sources for WSNs**

| Energy Source | Classification | Performance (power density) | Weakness | Strength |
|---|---|---|---|---|
| Solar Power | Radiant Energy | 100mW/cm$^3$ | Require exposure to light, and low efficiency if device is in building | Can use without limit |
| RF Waves | Radiant Energy | 0.02μW/cm$^2$ at 5Km from AM Radio | Low efficiency inside a building | Can use without limit |
| RF Energy | Radiant Energy | 40μW/cm$^2$ at 10m | Low efficiency if out of line of sight | Can use without limit |
| Body Heat | Thermal Energy | 60μW/cm$^2$ at 5$^o$C | Available only when temperature different is High | Easy to build using Thermocouple |
| External Heat | Thermal Energy | 135μW/cm$^2$ at 10$^o$C | Available only when temperature different is High | Easy to build using Thermocouple |

**Comparison of Energy Harvesting Sources for WSNs (Cont'd)**

| Body Motion | Mechanical Energy | 800μW/cm$^3$ | Dependent on Motion | High power density, not limited on interior and exterior |
|---|---|---|---|---|
| Blood Flow | Mechanical Energy | 0.93W at 100mmHg | Energy conversion efficiency is low | High power density, not limited on interior and exterior |
| Air Flow | Mechanical Energy | 177μW/cm$^3$ | Efficiency is low inside a building | High power density, |
| Vibrations | Mechanical Energy | 4μW/cm$^3$ | Has to exist at surrounding | High power density, not limited on interior and exterior |
| Piezoelectric | Mechanical Energy | 50μJ/N | Has to exist at surrounding | High power density, not limited on interior and exterior |

Beside the harvested energy for the Sensor network, the consumption of the harvested power for the different mode of the network has to be look upon before choosing power harvesting source. A review of some power consumption in some selected sensor nodes can be found in [23]. For some commercial sensor network nodes, the consumption differs, as shown in Table 2 below; power consumption of the nodes differs from manufacturers.

**Table 2.0 Comparison of Power consumption of some selected sensor network nodes**

| Operating conditions | Manufactures | | | |
|---|---|---|---|---|
| | Crossbow MICAz [24] | Waspmote [25-26] | Intel IMote2 [27] | Jennic JN5139 [28] |
| Radio standard | IEEE 802.15.4/Zigbee | IEEE 802.15.4/Zigbee | IEEE 802.15.4 | IEEE 802.15.4/Zigbee |
| Typical range | 100m (outdoor), 30m (indoor) | 500m | 30m | 1km |
| Data rate | 250 kbps | 250 kbps | 250 kbps | 250 kbps |
| Sleep mode (deep sleep) | 15 µA | 62 µA | 390 µA | 2.8 µA |
| Processor consumption | 8 mA active mode | 9 mA | 31-53 mA | 2.7+0.325 mA/MHz |
| Transmission | 17.4 mA (+0 dBm) | 50.26 mA | 44 mA | 34 mA (+3 dBm) |
| Reception | 19.7 mA | 49.56 mA | 44 mA | 34 mA |
| Supply voltage (Min) | 2.7 V | 3.3 V | 3.2 V | 2.7 V |
| Average power | 2.8 mW | 1 mW | 12 mW | 3 mW |

## 2.1 Ambient RF Sources and Available Power

A possible source of energy comes from ubiquitous radio transmitters. Radio waves, a part of electromagnetic spectrum consists of magnetic and electrical component. Radio waves carry information by varying a combination of the amplitude, frequency and phase of the wave within a frequency band. On contact with a conductor such as an antenna, the Electromagnetic (EM) radiation induces electrical current on the conductor's surface, known as skin effect. The communication devices uses antenna for transmission and/or reception of data by utilizing the different frequencies spectrum from 10Kz to 30Kz. The maximum theoretical power available for RF energy harvesting is 7.0μW and 1.0μW for 2.4GHz and 900MHz frequency respectively for free space distance of 40 meters. The path loss of signals will be different in environment other than free space [29]. Though for our work using the Powercast harvester, the power available for P2110 which operate at 915MHz is 3.5mW before conversion and 1.93mW after conversion at a distance of 0.6 meters, and 1μW at a distance of 11 meters [30].

**3.0    RF ENERGY HARVESTING AND THE USE OF POWERCAST HARVESTER**

RF power harvesting is a process whereby Radio frequency energy emitted by sources that generate high electromagnetic fields such as TV signals, wireless radio networks and cell phone towers, but through power generating circuit linked to a receiving antenna, captured and converted into usable DC voltage. Most commonly used as an application for radio frequency identification tags in which the sensing device wirelessly sends a radio frequency to a harvesting device which supplies just enough power to send back identification information specific to the item of interest. The circuit systems which receive the detected radio frequency from the antenna are made on a fraction of a micrometer scale but can convert the propagated electromagnetic waves to low voltage DC power at distances up to 100 meters. Depending on concentration levels which can differ through the day, the power conversion circuit may be attached to a capacitor which can disperse a constant required voltage for the sensor and circuit when there isn't a sufficient supply of incoming energy. Most circuits use a floating gate transistor as the diode which converts the signal into generated power but in linked to the drain of the transistor and a second floating gate transistor linked to a second capacitor can enable a higher output voltage once the capacitors reach full potential [31].

Though the effectiveness of energy harvesting depends largely on the amount and predictable availability of energy source; whether from radio waves, thermal differentials, solar or light sources, or even vibration sources. There are three categories for ambient energy availability: intentional, anticipated, and unknown as shown in Figure 1 below

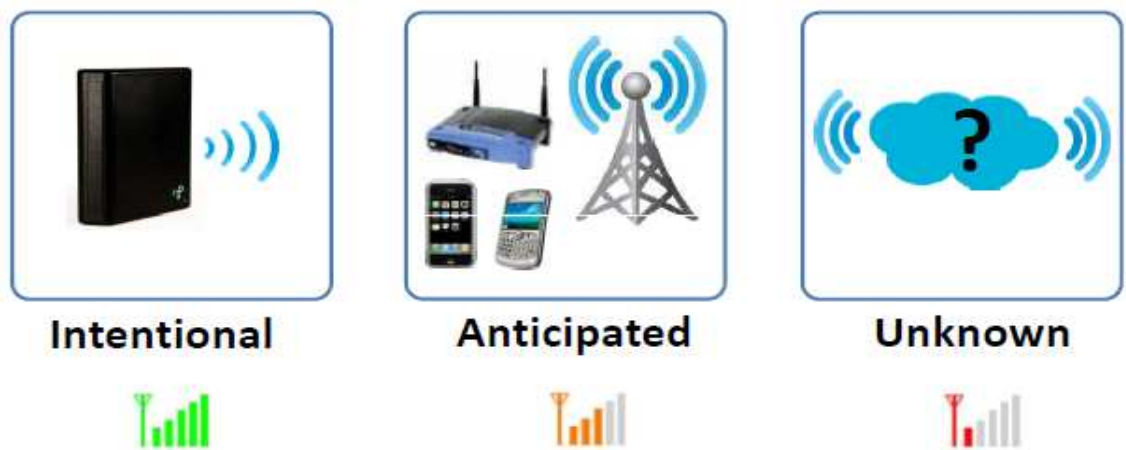

Fig.1. Pictorial view of intentional, anticipated, and unknown energy sources

Our research relies basically on the intentional using the Powercast harvester.

**3.1    Intentional Energy Harvesting:** The designs rely on an active component in the system, such as an RF transmitter that can explicitly provide the desired type of energy into the environment when the device needs it. Powercast support this approach with an energy source of 3W 915MHz RF transmitters, the P1110 and P2110 also use along with it as receiver. The intentional energy approach is also appropriate for other types of energy, such as placing an energy harvesting on a piece of industrial equipment that vibrates when it is operating. Using an intentional energy source allows designers to engineers a consistent energy solution. A quick look at the basic operation of the Transmitter and receiving circuit is as discussed below.

## 3.2 The Powercast TX91501 Powercaster Transmitter

The Powercast TX91501 is a radio frequency power transmitter specifically designed to provide both power and data to end devices containing the Powercast P2110 or P1110 power harvester receivers [30]. The transmitter is housed in a durable plastic case with mounting holes. It is powered by a regulated 5V DC voltage mostly from a power source of 240V AC, rectified and regulated to its accommodated voltage of 5V DC from its in-built internal circuitry. The transmitter has a factory set, fixed power output and no user adjustable settings. Also a beautiful and control feature of it is the status LED which provide a feedback on functional state. It provides a maximum of 3Watts EIRP (Equivalent or Effective Isotropic Radiated Power). A side view, real view and its transmission state are as shown below in Fig2.

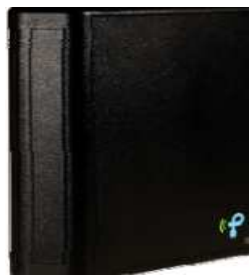 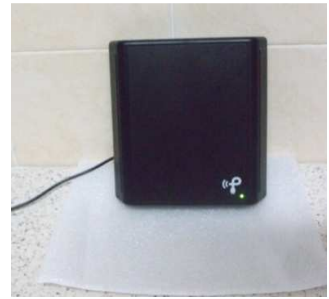

Fig2 (a), a side view, and (b) a real view of a TX91501 Powercaster Transmitter in its transmission state

The Powercast transmitter transmits power in the form of Direct Sequence Spread Spectrum (DSSS) and Data in the Amplitude Shift Keying (ASK) modulation and at a center frequency of 915MHz. The power output is 3 watts EIRP and vertically polarizes for optimal transmission.

For data communication, it has an 8-bit factory set, TX91501 identification (ID) number broadcast with random intervals up to 10ms using Amplitude Shift Keying (ASK) modulation. Its operating temperature is within the range of -20oC to 50oC at the power input from mains of 5VDC/1A.

## 3.3 The Powercast Power Harvester Receiver

The Powercast Receivers can harvest directly a directed or ambient RF energy and convert it to DC power for remotely recharging batteries or battery free devices. The two modules available for our research are P1110 and P2110 and both have similarities and differences in their area of applications as shown in table 3.0 below.

**Table 3.0 Comparison of the Two RF Energy Powercast receivers**

| Receivers | Differences | Similarities |
|---|---|---|
| P2110 | 1. Design for battery charging and direct power applications. <br> 2. Provide intermittent/pulsed power output <br> 3. Configurable, regulated output voltage up to 5.25V <br> 4. Power management and control I/O for system optimization | 1. Harvesting range from 850-950 MHz <br> 2. Works with standard 50-ohm antennas |
| P1110 | 1. Design for battery charging and direct power applications <br> 2. Configurable over voltage protection up to 4.2V <br> 3. Connect directly to rechargeable batteries including Alkaline, Lithium Ion, and Ni-MH. | 1. Harvesting range from 850-950 MHz <br> 2. Works with standard 50-ohm antennas |

The Powercast P2110 power harvester receiver is an RF energy harvesting device that converts RF energy to DC voltage. It has wide RF operating range, provides RF energy harvesting and power management for battery free, micro power devices. It converts the RF energy to DC and stores in a capacitor as well as boosting the voltage to the set output voltage level and enables the voltage output.

### 3.4 Measurement of RF Power Received and Gains

Power meters which provides most accurate measurement of RF power of any of the types of RF measurement equipments, and the simplified Friis Equation that provides a reasonable estimate of the amount of power that is received and available for use were utilized in our experiment.

#### 3.4.1 Friis Transmission Equation

Friis Transmission Formula is sorely used to study RF communication links [32]. The formula can be used in situations where the distance between two antennas is known, and a suitable antenna need to be found. Using Friis transmission equation, one can solve for the antenna gains needed at either the transmitter or receiver in order to meet certain design specifications.

$$\frac{P_r}{P_t} = G_t G_r \left(\frac{\lambda}{4\pi R}\right)^2 \qquad (1)$$

Where $P_r$ is the received power in Watts (W), $P_t$ is the transmitted power, $G_t$ is the transmitting antenna's gain, $G_r$ is the receiving antenna's gain, $\lambda$ is the wavelength of the transmitted and received signal in meters, and R is the distance between the antennas in meters. The gain of the antennas, usually measured in decibels, can be converted to power ratio using;

$$G = 10^{\frac{G_{dB}}{10}} \qquad (2)$$

$$\lambda = \frac{c}{f} \tag{3}$$

Where C is the speed of light in meters per second, and f is the frequency in Hz. Hence, C is equal to 3 x $10^8$ m/s

A simplified version of the Friis equation [33] is provided by the Powercast company for quick and easiest calculation on a spread sheet, where a reasonable estimate of the amount of power generated, received and available for use are calculated.

### 3.4.2 Power Density

Radio Frequency (RF) propagation is defined as the travel of electromagnetic waves through or along a medium. For RF propagation between approximately 100 MHz and 10 GHz, radio waves travel very much as they do in free space and travel in a direct line of sight and a slight difference in the dielectric constants of air and space [34]. For air, the dielectric is one and 1.000536 at sea level. In Antennas theory, an isotropic radiator is a theoretical, lossless, Omni-directional (spherical) antenna [34-35]. That is, it radiates uniformly in all directions. The power of a transmitter that is radiated from an isotropic antenna will have a uniform power density (power per unit area) in all directions. Power density at any distance from an isotropic antenna is the ratio of the transmitted power by the surface area of a sphere ($4\pi R^2$) at that distance. The surface area of the sphere increases by the square of the radius, therefore the power density, $P_D$, (watts/square meter) decreases by the square of the radius

$$P_D = \frac{P_t}{4\pi R^2} \tag{4}$$

Where, $P_t$ is the peak or average power, $P_D$ the power density and R the distance between the transmitter and the receiving antenna. Radars use directional antennas to channel most of the radiated power in a particular direction. The Gain ($G_t$) of an antenna is the ratio of power radiated in the desired direction as compared to the power radiated from an isotropic antenna, or:

$$G_t = \frac{\text{Maximum radiation intensity of actual antenna}}{\text{Radiation intensity of isotropic antenna with same power input}} \quad (5)$$

The power density at a distant point from radar with an antenna gain of $G_t$ is the power density from an isotropic antenna multiplied by the radar antenna gain. Power density from radar,

$$P_D = \frac{P_t \, G_t}{4\pi R^2} \quad (6)$$

### 3.5 Energy Storage

The most common energy storage device used in a sensor node is a battery, either non rechargeable or rechargeable. A non-rechargeable battery (e.g., alkaline) is suitable for a micro sensor with very low power consumption (e.g., 50 µW). Alternatively, a rechargeable battery (e.g., lithium ion) is used widely in sensor nodes with energy harvesting technology [36]. A battery is used not only for storage of energy generated by the harvesting device but also to regulate the supply of energy to a sensor node. As wireless sensor node is powered by exhaustible batteries [37]. Several factors affect the quality of these batteries, but the main factor is cost. In a large-scale deployment, the cost of hundreds and thousands of batteries is a serious deployment constraint. Batteries are specified by a rated current capacity, *C*, expressed in ampere-hour. This quantity describes the rate at which a battery discharges without significantly affecting the prescribed supply voltage (or potential difference). Practically, as the discharge rate increases, the rated capacity decreases. Most portable batteries are rated at $1C$. This means a 1000mAh battery provides 1000mA for 1 hour, if it is discharged at a rate of $1C$. Ideally, the same battery can discharge at a rate of $0.5C$, providing 500mA for 2 hours; and at $2C$, 2000mA for 30 minutes and so on. $1C$ is often referred to as a 1-hour discharge. Likewise, a $0.5C$ would be a 2-hour and a $0.1C$ a 10-hour discharge. In reality, batteries perform at less than the prescribed rate. Often, the Peukert Equation is applied to quantifying the capacity offset (i.e., how long a battery lasts in reality):

$$T = \frac{C}{I^n} \tag{7}$$

where C is the theoretical capacity of the battery expressed in ampere-hours; I is the current drawn in Ampere (A); T is the time of discharge in seconds, and n is the Peukert number, a constant that directly relates to the internal resistance of the battery. The value of the Peukert number indicates how well a battery performs under continuous heavy current. A value close to 1 indicates that the battery performs well; the higher the number, the more capacity is lost when the battery is discharged at high current. The Peukert number of a battery is determined empirically. For example, for lead acid batteries, the number is typically between 1.3, and 1.4. Drawing current at a rate greater than the discharge rate results in a current consumption rate higher than the rate of diffusion of the active elements in the electrolyte. If this process continues for a long time, the electrodes run out of active material even though the electrolyte has not yet exhausted its active materials. This situation can be overcome by intermittently drawing current from the battery and also proper power management techniques.

## 4.0 ENERGY MANAGEMENT IN WSNs

Despite the fact that energy scavenging mechanisms can be adopted to recharge batteries, e.g., through Powercast harvesters [30], solar panels [2], piezoelectric or acoustic transducers [21], energy is a limited resource and must be used judiciously. Hence, efficient energy management strategies must be devised at the sensor nodes to prolong the network lifetime as much as possible. Many routing, power management, and data dissemination protocols have been specially designed for WSNs [38]. The EAGRP [39], An Enhanced AODV [40], An Energy-Efficient Ant Based Routing Algorithm for Wireless Sensor Networks [41], all have developed different protocols in order to manage the available energy in WSNs. In a related work [42], use Energy-hungry Sensors in trying to manage the available energy in WSNs. Reducing power consumption has become a major challenge in wireless sensor networks. As a vital factor affecting system cost and lifetime, energy consumption in wireless sensor networks is an emerging and active research area. The energy consumption of WSNs is of crucial concern due to the limited availability of energy. Whereas energy is a scarce resource in every wireless device, the problem in WSNs is more severe for the following reasons [37]:

1. Compared to the complexity of the task they carry out; sensing, processing, self-managing, and communication, the nodes been very small in size to accommodate high-capacity power supplies.

2. While the research community is investigating the contribution of renewable energy and self recharging mechanisms, the size of nodes is still a constraining factor.

3. Ideally, a WSN consists of a large number of nodes. This makes manually changing, replacing or recharging batteries almost impossible.

4. The failure of a few nodes may cause the entire network to fragment prematurely. The problem of power consumption can be approached from two angles. One is to develop energy efficient communication protocols (self-organization, medium access, and routing protocols) that take the peculiarities of WSNs into account. The other is to identify activities in the networks that are both wasteful and unnecessary and mitigate their impact. Wasteful and unnecessary activities can be described as local (limited to a node) or global (having a scope network-wide). In either case, these activities can be further considered as accidental side-effects or results of non optimal software and hardware implementations (configurations). For example, observations based on field deployment reveal that some nodes exhausted their batteries prematurely because of unexpected overhearing of traffic that caused the communication subsystem to become operational for a longer time than originally intended [43]. Similarly, some nodes exhausted their batteries prematurely because they aimlessly attempted to establish links with a network that had become no longer accessible to them. Most inefficient activities are, however, results of non optimal configurations in hardware and software components. For example, a considerable amount of energy is wasted by an idle processing or a communication subsystem. A radio that aimlessly senses the media or overhears while neighboring nodes communicate with each other consumes a significant amount of power. A dynamic power management (DPM) control strategy aimed at adapting the power/performance of a system to its workload. The DPM having a local or global scope, or both aims at minimizing power consumption of individual nodes by providing each subsystem with the amount of power that is sufficient to carry out a task at hand [37]. Hence, it does not consider the residual energy of neighboring nodes. IEEABR as the proposed algorithm considers the available power of nodes and the energy consumption of each path as the reliance of routing selection. It improves memory usage and utilizes the self organization, self-adaptability, and dynamic optimization capability of the ant colony system to find the optimal path and multiple candidate paths from source nodes to sink node. The protocol avoiding using up the energy of nodes on the optimal path and prolongs the network lifetime while preserving network connectivity. This is necessary since for any WSN protocol design, the important issue is the energy efficiency of the underlying algorithm due to the fact that the network under investigation has strict power requirements. It has been proposed [44], for forward ants sent directly to the sink-node; the routing tables only need to save the neighbor nodes that are in the direction of the sink-node. This considerably reduces the size of the routing tables and, in consequence, the memory needed by the nodes. Since one of the main concerns in WSN is to maximize the lifetime of the network, which means saving as much energy as possible, it would be preferable that the routing algorithm could perform as much processing as possible in the network nodes, than transmitting all data through the ants to the sink-node to be processed there. In fact, in huge sensor networks where the number of nodes can easily reach more than Thousands of units, the memory of the ants would be so large that it would be unfeasible to send the ants through the network. To implement these ideas, the memory $M_k$ of each ant is reduced to just two records, the last two visited nodes [41]. Since the path followed by the ants is no more in their memories, a memory must be created at each node that keeps record of each ant that was received and sent. Each memory record saves the previous node, the forward node, the ant identification and a timeout value. Whenever a forward ant is received, the node looks into its memory and searches the ant identification for a possible loop. If no record is found, the node saves the required information, restarts a timer, and forwards the ant to the next node. If a record containing the ant identification is found, the ant is eliminated. When a node receives a backward ant, it searches its memory to find the next node to where the ant must be sent. In this section, we modify the EEABR to improve the Energy consumption in the nodes of WSNs and also to in turn improve the performance and efficiency of the networks. The main focus of this chapter is on IEEABR power management strategies in WSNs.

The Algorithm of our proposed power management techniques is as below.

1. Initialize the routing tables with a uniform probability distribution;

$$P_{ld} = \frac{1}{N_k} \tag{8}$$

Where $P_{ld}$ is the probability of jumping from node l to node d (destination), $N_k$ the number of nodes.

2. At regular intervals, from every network node, a forward ant k is launched with the aim to find a path until the destination. The identifier of every visited node is saved onto a memory $M_k$ and carried by the ant.

Let k be any network node; its routing table will have N entries, one for each possible destination.

Let d be one entry of k routing table (a possible destination).

Let $N_k$ be set of neighboring nodes of node k.

Let $P_{kl}$ be the probability with which an ant or data packet in k, jumps to a node l, l∈$N_k$, when the destination is d ($d \neq k$). Then, for each of the N entries in the node k routing table, it will be $n_k$ values of $P_{ld}$ subject to the condition:

$$\sum_{l \in N_k} P_{ld} = 1; \quad d = 1, \dots, N \tag{9}$$

3. At every visited node, a forward ant assigns a greater probability to a destination node d for which falls to be the destination among the neighbor node, d ∈ $N_k$. Hence, initial probability in the routing table of k is then:

$$P_{dd} = \frac{9N_k - 5}{4N_k^2} \qquad (10)$$

Also, for the rest neighboring nodes among the neighbors for which m ∈ $N_k$ will then be:

$$P_{dm} = \begin{cases} \frac{4N_k - 5}{4N_k^2}, & if\ N_k > 1 \\ 0, & if\ N_k = 1 \end{cases} \qquad (11)$$

Of course equation (10) and (11) satisfy (9). But if it falls to the case where by none among the neighbor is a destination, equation (8) applies to all the neighboring nodes.

Else,

4. Forward ant selects the next hop node using the same probabilistic rule proposed in the ACO metaheuristic:

$$P_k(r,s) = \begin{cases} \frac{[\tau(r,s)]^\alpha \cdot [E(s)]^\beta}{\sum_{u \notin M_k} [\tau(r,u)]^\alpha \cdot [E(s)]^\beta}, & s \notin M_k \\ 0, else \end{cases} \qquad (12)$$

Where $p_k(r,s)$ is the probability with which ant k chooses to move from node r to node s, $\tau$ is the routing table at each node that stores the amount of pheromone trail on connection (r,s), E is the visibility function given by $\frac{1}{(C - e_s)}$ (c is the initial energy level of the nodes and $e_s$ is the actual energy level of node s), and α and β are parameters that control the relative importance of trail versus visibility. The selection probability is a trade-off between visibility (which says that nodes with more energy should be chosen with high probability) and actual trail intensity (that says that if on connection (r,s) there has been a lot of traffic then it is highly desirable to use that connection.

5. When a forward ant reaches the destination node, it is transformed in a backward ant which mission is now to update the pheromone trail of the path it used to reach the destination and that is stored in its memory.

6. Before backward ant k starts its return journey, the destination node computes the amount of pheromone trail that the ant will drop during its journey:

$$\Delta \tau = \frac{1}{C - \left[\frac{EMin_k - Fd_k}{EAvg_k - Fd_k}\right]} \tag{13}$$

And the equation used to update the routing tables at each node is:

$$\tau(r,s) = (1-\rho) * \tau(r,s) + \left[\frac{\Delta \tau}{\phi . Bd_k}\right] \tag{14}$$

Where $\phi$ a coefficient and $Bd_k$ is the distance travelled (the number of visited nodes) by the backward ant k until node r. which the two parameters will force the ant to lose part of the pheromone strength during its way to the source node. The idea behind the behavior is to build a better pheromone distribution (nodes near the sink node will have more pheromone levels) and will force remote nodes to find better paths. Such behavior is important when the sink node is able to move, since pheromone adaptation will be much quicker [41].

7. When the backward ant reaches the node where it was created, its mission is finished and the ant is eliminated.

By performing this algorithm several iterations, each node will be able to know which are the best neighbors (in terms of the optimal function represented by Equation (14)) to send a packet towards a specific destination. The flow chart describing the action of movement of forward ant for our proposed Algorithm is as shown below in Figure 3.0. The backward ant takes the opposite direction of the flow chart.

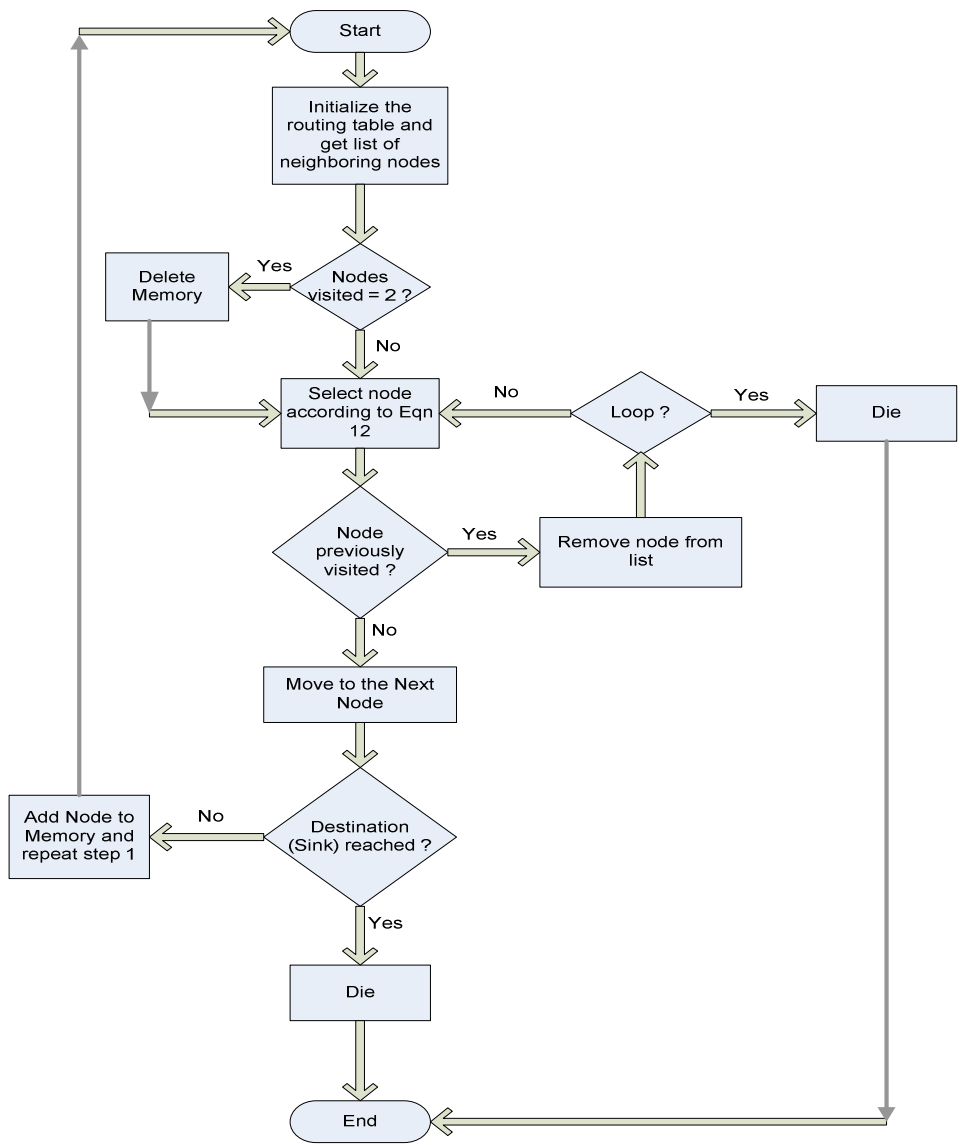

Fig 3.0 An IEEABR forward ant flow chart

## 4.1 Algorithm Operations

After the initialization of the routing table, and setting up a forward ant for hoping from node to node in search for the sink, at every point in time, a node becomes a source holding in its stack or memory information about an event around itself (neighbors). The information gathered in its memory is transferred or disseminated towards the sink node with the help of neighbor nodes behaving as repeaters. Associated raw data generated at each source (nodes) is divided in to M pieces known as data parts. An integer value M also represents the number of ant agents involving in each routing task. This raw data provided by the source node about an event contains information such as source node identification, event identification, time and the data about the event. The data size is chosen based on the sensor nodes deployed and the size of the buffer. After the splitting of the raw data, each part is associated with routing parameters to build a data packet ready to transfer. These parameters are code identification, describing the code following as data, error or acknowledge; $C_{ID}$. Next is the node identification to which the packet is transferred; $N_{ID}$. Packet number also represents the ant agent k; $S_N$ is the sequence number, $N_k$ which contains the number of visited nodes so far, and the $k^{th}$ data part as shown in Figure 4.0 below. In this figure, the group of the first four fields is named the data header. When delivery of all data packages is accomplished, the base combines them into raw data.

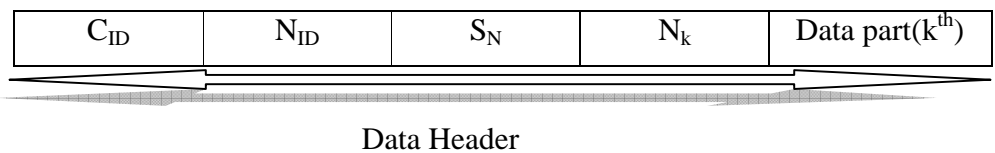

Data Header

Fig 4.0 Data packet content

When a node participating in a routing received a data packet whose agent number is given, it makes decision about the next destination for that packet of data. The decision on the next node or destination for which the packet of data should be transfer, will depend on equation (10) else (12) with the highest $P_k(r,s)$. The pheromone level of the neighbors which is the first determining factor follows by the Energy levels of the neighbor nodes which are most important on the decision rule. For any of the neighbor chosen, the $N_{ID}$ field of the node is updated and the packet is then broadcasted. The remaining neighbors among the chosen node also hear the broadcast, they check the $N_{ID}$ field and understand that the message is not made for them; they as such quickly discard the packet immediately after only listening to the $N_{ID}$ field of the packet. The $N_k$ is updated with an increment by one after ensuring that $S_N$ is not in the list of the tabus (routing table) of the chosen node. The next node is determined to update the $N_{ID}$ field by performing the same operation as perform earlier by the first node and the sequence continue till the packet gets to the sink node. The reversed operation is done for the backward ant as for the acknowledgement, which get to the source now the last bus stop for the backward ant and die off after reaching the source.

## 5.0 EXPERIMENT AND SIMULATION RESULTS

Different experiments conducted to measure the circuit's parameter and influence of the RF power source, Simulation results based on the performance of the circuit with differences in distance of the harvester from the power sources, the energy usage, and energy management using our proposed IEEABR, are all analyze below while also, showing the harvesting set-up and the simulation environment.

### 5.1     Experimental Results

Using the Powercast Calculator, and setting components; P2110 at 1.2V-915 MHz, Battery Capacity at 1150 mAh for P2110 and P1110 at 4.0V-915 MHz for the same battery capacity, while varying the distance between the transmitter, the readings are as shown in Table 4a, b, c and d with differences in the receiver and antenna used in the experiment. The behavior of the packets received with time is shown in Fig 5a, and b, while for the packet received with distance for different harvesters and antennas are compared in figure 6a and b.

**Table 4.0 (a) Amount of Power harvested by P2110 harvester using Dipole Antenna**

| Distance (ft) | P (µW) | I (µA) | Recharge Time (hrs) |
|---|---|---|---|
| 2 | 3687 | 3073 | 22.08 |
| 5 | 523 | 436 | 155.04 |
| 10 | 135 | 112 | 602.64 |
| 12 | 85 | 71 | 952.32 |
| 15 | 37 | 31 | 2169.12 |
| 18 | 11 | 9 | 7360.56 |
| 20 | 1 | 1 | 68339.28 |

**Table 4.0 (b) Amount of Power harvested by P2110 harvester using Patch Antenna**

| Distance (ft) | P (µW) | I (µA) | Recharge Time (hrs) |
|---|---|---|---|
| 5 | 1925 | 1604 | 42.24 |
| 10 | 386 | 322 | 210.50 |
| 15 | 189 | 158 | 429.40 |
| 18 | 131 | 109 | 618.5 |
| 20 | 102 | 85 | 797.50 |
| 25 | 50 | 41 | 1639.00 |
| 30 | 19 | 16 | 4353.00 |
| 35 | 5 | 4 | 15517.00 |
| 36 | 1 | 1 | 70019.00 |

**Table 4.0 (c) Amount of Power harvested by P1110 harvester using Dipole Antenna**

| Distance (ft) | P (µW) | I (µA) | Recharge Time (hrs) |
|---|---|---|---|
| 2 | 3688 | 922 | 62.40 |
| 4 | 1085 | 271 | 211.92 |
| 6 | 259 | 65 | 888.72 |
| 7 | 86 | 22 | 2659.92 |

**Table 4.0 (d) Amount of Power harvested by P1110 harvester using Patch Antenna**

| Distance (ft) | P (µW) | I (µA) | Recharge Time (hrs) |
|---|---|---|---|
| 2 | 16115 | 4029 | 14.16 |
| 4 | 3070 | 768 | 74.88 |
| 6 | 1551 | 388 | 148.30 |
| 8 | 810 | 203 | 283.90 |
| 10 | 366 | 92 | 627.60 |
| 12 | 93 | 23 | 2475.00 |
| 13 | 26 | 7 | 8750.00 |

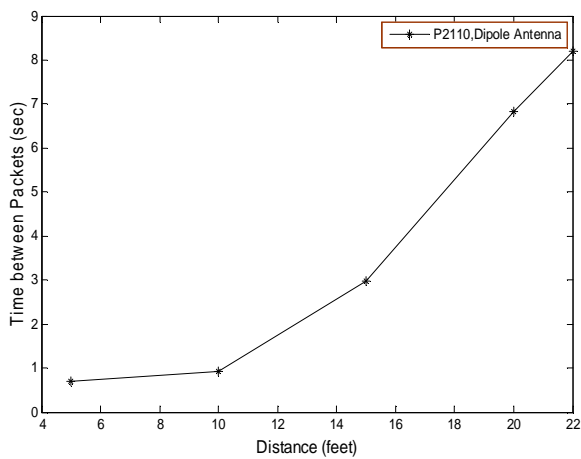
(a) TX91501-3W EIRP, 915MHz Power Transmitter

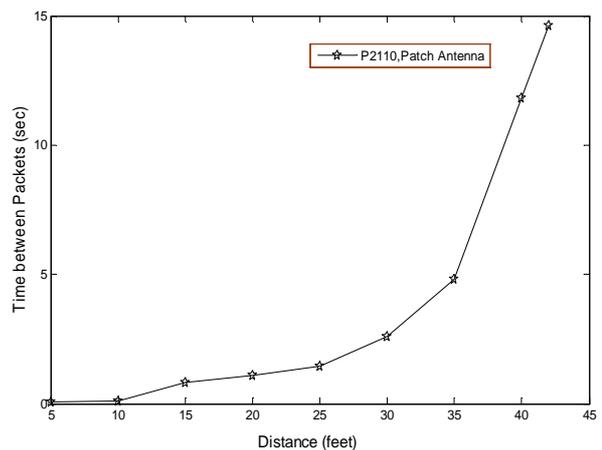
(b) TX91501-3W EIRP, 915MHz Power Transmitter

Fig 5.0 Variation in Time between Packets received and Distance of harvesting

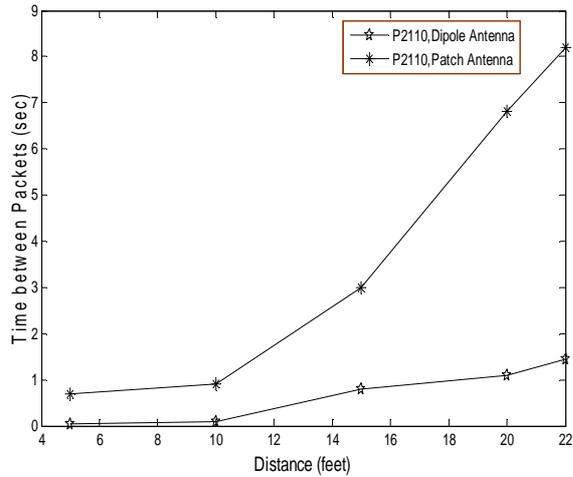 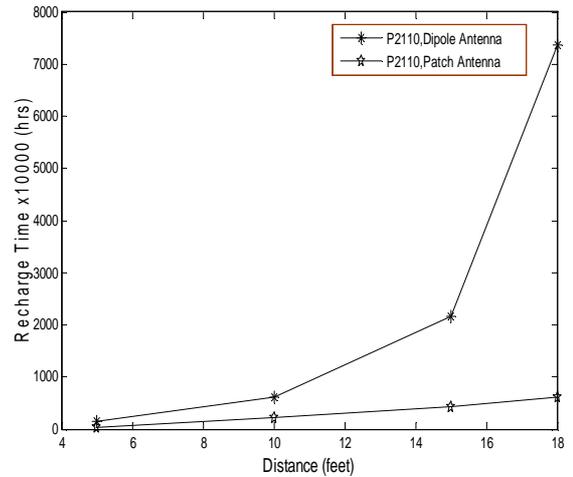

(a) Comparison of packet received with distance using different Antenna    (b) Comparison of Battery Recharge Time, using Dipole and Patch Antennas with P2110

Fig 6.0 Comparison of Power harvesting using Dipole and Patch Antenna with P2110

## 5.2    Simulation Results

We use event driven network simulator-2 (NS-2) [45] based on the network topology to be able to evaluate the implementation of the proposed Energy Management Protocol. This software provides a high simulation environment for wireless communication with detailed propagation, MAC and radio layers. AntSense (an NS-2 module for Ant Colony Optimization) [46] was used for the EEABR. The simulation parameters are as shown in table 5.0. We assume that all nodes have no mobility since the nodes are fixed in application of most wireless sensor networks. Simulations were run for 60 minutes (3600 seconds) each time the simulation starts, and the remaining energy of all nodes were taken and recorded at the end of each simulation. The average energy calculated while also noting the minimum energy of the nodes. This helps in taking tracks of the performance of the management protocols in term of network's energy consumption.

**Table 5.0 Simulation Parameters**

| Parameters | Values |
|---|---|
| Routing Protocols | AODV, EEABR, IEEABR |
| MAC Layer | IEEE 802.15.4 |
| Frequency | 2.4GHz |
| Packet Size | 1 Mb |
| Area of Deployment | 200x200 $m^2$ (10 nodes), 300x300 $m^2$ (20 nodes), 400x400 $m^2$ (30 nodes), 500x500 $m^2$ (40 nodes), 600x600 $m^2$ (50-100 nodes), |
| Data Traffic | Constant Bit Rate (CBR) |
| Simulation Time | 3600 sec. |
| Battery power | 1150mAH, 3.7V |
| Propagation Model | Two-ray ground reflection |
| Data rate | 250 Kbps |
| Current draw in Sleep Mode | 62μA |
| Current draw in Transmitting Mode | 50.26mA |
| Current draw in Receiving Mode | 49.56mA |
| Current draw in Idle Mode (Processor) | 9mA |

As Energy is the key parameter to be considered when designing protocol for power management to enhance maximum life time of sensor networks, we therefore use 1. The Minimum Energy which gives the lowest energy amount of all nodes at the end of simulations, 2. The Average Energy which represents the average of energy of all nodes at the end of simulation, and 3. The simulation was done on a static WSN, where sensor nodes were randomly deployed with objective to monitor a static environment. Nodes were responsible to monitor and send the relevant sensor data to the sink node in which nodes near the phenomenon will depreciate easily in energy as they will be forced to periodically transmit data. Simulations were run for 60 minutes (3600 seconds) each time the simulation starts, and the remaining energy of all nodes were taken and recorded at the end of each simulation. The average energy calculated while also noting the minimum energy of the nodes. Figure 8, presents the results of the simulation for the studied parameters; The Average Energy, and Minimum Energy of AODV, EEABR and IEEABR. As it can be seen from the results presented in the figures 8.0 below, the IEEABR protocol had better results in both Average energy of the nodes and the Minimum energy of node experienced at the end of simulation. The AODV as compared to the EEABR perform worst in all cases. In term of Average energy levels of the network, IEEABR as compared with EEABR average energy values, varies between 2% and 8%, while for AODV is in the range of 15% to 22% also to the minimum energy of the nodes. Fig. 7.0 below shows a screenshot of a NAM window of the simulation environment for 10 nodes randomly deployed, while the results of the simulations are as shown in Fig.8.0

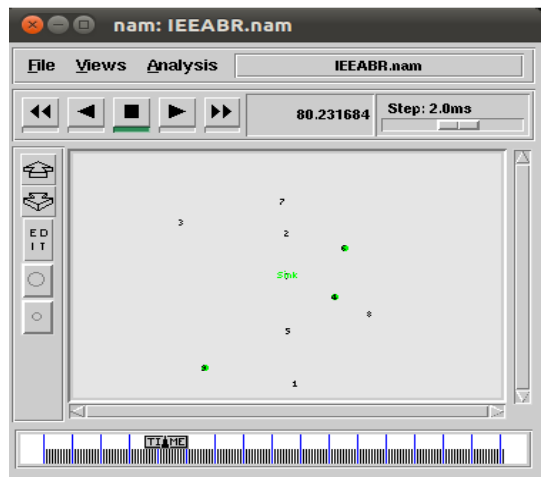

Fig. 7.0 Graphical Representation of the Simulation Environment in NS-2.34 with 10 Nodes

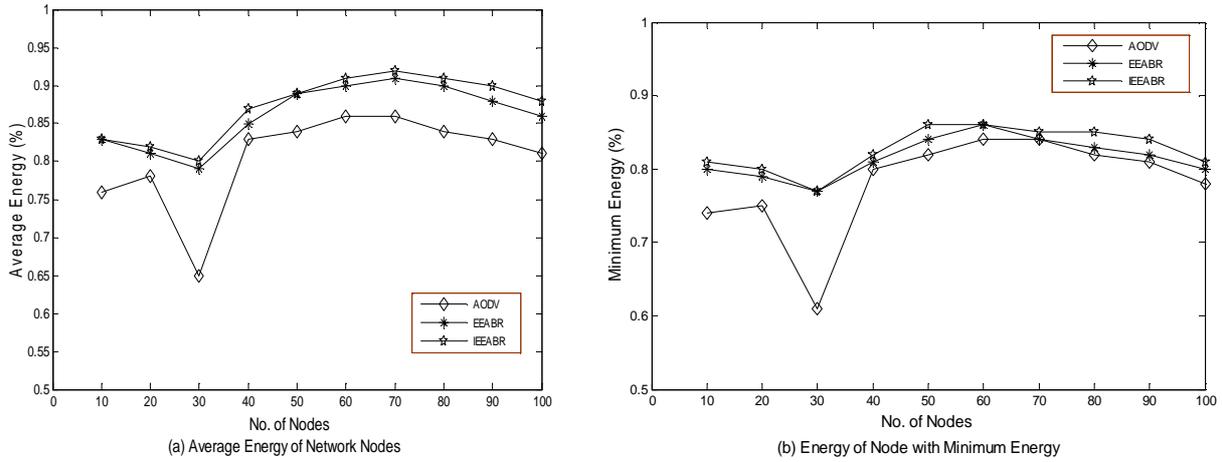

Fig. 8.0: Performance analysis of AODV, EEABR and IEEABR Energy Management Protocols.

## 5.3 Real-Time Implementation of RF Powercast Energy Harvester

The real-time implementation in Harvesting Radio Frequency Energy from Powercast Harvesters and the management of the power harvested using the Improved Energy Efficient Ant Based Routing Algorithm is presented as shown in Figure 9 below. The experiment was set up and measurement of available power measured while also varying the distance between the harvester and transmitter. The time between each packet delivery that is, the harvesting period where noticed and recorded. The power consumption of the Waspmote under consideration can be found in [25-26]. The battery powering the Waspmote is 1150mAh at 3.7V, which can sufficiently power each of the nodes separately under constant transmission or reception for 19.39 hrs. For our management protocol applied, the maximum energy consumed was found to be 23% of the supply energy amounting to total current drawn to be 264.5mA for 1 hr. It then means that, the battery can sufficiently powered the node with the minimum energy for 4.35 hrs without recharging. For the recharging of the battery at 15 feet as shown in table 4b, it takes 429.4 hrs for fully recharge when empty, and 91.9 hrs to replenish the drawn current of 264.5mA. But with constant harvesting, it then means that the total energy of the battery remains without reduction, which can then sustain the network for the required number of years needed for sensing. A quick look at the receiver in its receiving and conversion state with Dipole, and Patch antenna connected for the application of harvesting from the Powercast transmitter and the harvesting mode of the receiver 3 feet (0.914m) away from the transmitter respectively are shown in Fig 9(a), (b), and (c). P2110 Powercast harvester Receiver with (a) Dipole (Omni-directional) antenna, (b) Patch (directional) antenna, (c) TX95101 Powercaster transmitter in its harvesting mode, (d) (a) with Waspmote, and (e) Gateway connected to the Sink. The results of the measurements of the harvested RF energy are as presented in table 4.0 (a-d) and Fig. 5.0 (a-b).

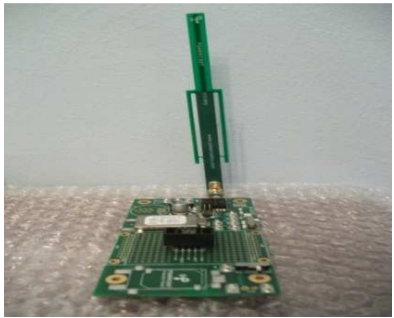 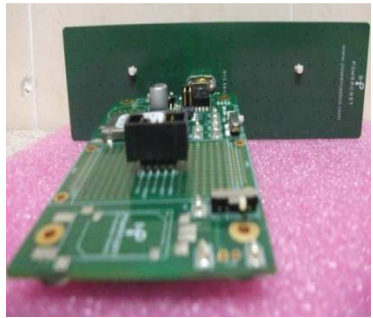 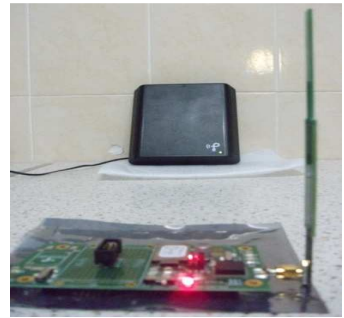

(a), (b), (c)

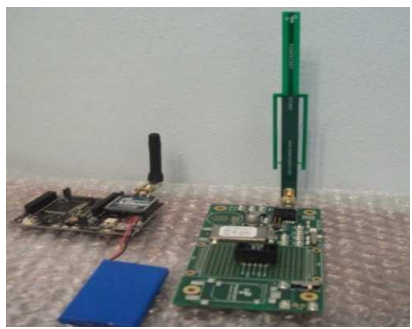 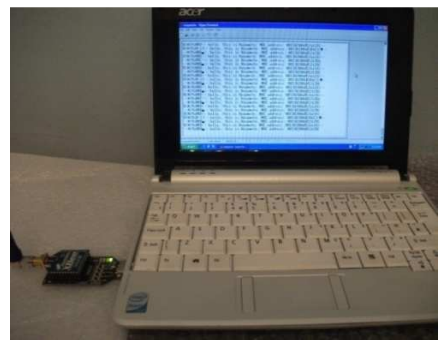

(d), (e)

Fig. 9.0 Hardware set-up of the real-time implementation

## 6.0    CONCLUSIONS AND FUTURE WORK:

In this chapter, research based on the application of Radio Frequency Energy harvesting, using Powercast harvesters to support the limited available energy of wireless sensor networks, and its management using Ant Colony Optimization metaheuristic was adopted. In this work we proposed an Improved Energy Efficient Ant Based routing Algorithm energy management technique, which improves the lifetime of sensor networks. The IEEABR utilizes initialization of uniform probabilities distribution in the routing table while given special consideration to neighboring nodes which falls to be the destination (sinks) in other to save time in searching for the sink leading to reduced energy consumption by the nodes. The experimental results showed that the algorithm leads to very good results in different WSNs. Also looking at the harvested energy, the time of charging the battery powering the sensor nodes drastically reduced, while requiring time intervals of 91.9 hrs to recharge the battery. The protocol considers the residual energy of nodes in the network after each simulation period. Based on NS-2 simulation, the IEEABR approach has effectively balances the WSN node power consumption and increase the network lifetime. Consequently, our proposed algorithm can efficiently extend the network lifetime without performance degradation. This Algorithm focused mainly on energy management and the lifetime of wireless sensor networks.

As future work, we intend to build a linking circuit so as to directly charge the Waspmote battery from the Powercast harvesters, harvest the useless energy from the Waspmote, study a dual approach in the selection of sink, self destruction of the backward ants should there exist a link failure and an alternate means of retrieving the information carried by the backward ant to avoid lost of information. We also intend to design a Maximum Power Point Tracker (MPPT), so as to dual power the waspmote, and model both sources for perpetually operation of the sensor networks.

**AUTHORS**

1. **Adamu Murtala Zungeru**

   **PhD Research Student**, School of Electrical and Electronic Engineering, Faculty of Engineering, The University of Nottingham Malaysia campus,

   Jalan Broga, 43500 Semenyih, Selangor Darul Ehsan, Malaysia

   Email: keyx1mzd@nottingham.edu.my

2. **Dr. Li-Minn ANG**

   **Associate Professor**, School of Electrical and Electronic Engineering, Faculty of Engineering, The University of Nottingham Malaysia campus

   Email: kenneth.ang@nottingham.edu.my

3. **Professor. Dr. SRS Prabaharan**

   **Professor**, School of Electrical and Electronic Engineering, Faculty of Engineering, The University of Nottingham Malaysia campus

   Email: prabaharan.sahaya@nottingham.edu.my

4. **Dr. Kah Phooi Seng**

   **Associate Professor**, School of Electrical and Electronic Engineering, Faculty of Engineering, The University of Nottingham Malaysia campus

   Email: jasmine.seng@nottingham.edu.my